\documentclass{aip-cp}

\usepackage[numbers]{natbib}
\usepackage{rotating}
\usepackage{graphicx}
\usepackage{textcomp}

\begin{document}

\title{Actuator Development at IAAT for the Cherenkov Telescope Array Medium Size Telescopes}

\author[aff1]{S.~Diebold\corref{cor1}}
\author[aff1]{\unskip, J.~Dick}
\author[aff1]{\unskip, G.~P\"{u}hlhofer}
\author[aff1]{\unskip, S.~Renner}
\author[aff1]{\unskip, A.~Santangelo}
\author[aff1]{\unskip, T.~Schanz}
\author[aff1]{\unskip, C.~Tenzer}
\author{\unskip, and the CTA Consortium\corref{cor2}}

\affil[aff1]{Institut f{\"u}r Astronomie und Astrophysik, Abteilung Hochenergieastrophysik, Kepler Center for Astro and Particle Physics, Eberhard Karls Universit{\"a}t T{\"u}bingen, Sand 1, 72076 T{\"u}bingen, Germany}
\corresp[cor1]{Corresponding author: diebold@astro.uni-tuebingen.de}
\corresp[cor2]{See http://www.cta-observatory.org for the full author and affiliation list}

\maketitle

\begin{abstract}
The Cherenkov Telescope Array (CTA) will be the future observatory for TeV gamma-ray astronomy. In order to increase the sensitivity and to extend the energy coverage beyond the capabilities of current facilities, its design concept features telescopes of three different size classes. Based on the experience from H.E.S.S. phase II, the Institute for Astronomy and Astrophysics T\"{u}bingen (IAAT) develops actuators for the mirror control system of the CTA Medium Size Telescopes (MSTs). The goals of this effort are durability, high precision, and mechanical stability under all environmental conditions. Up to now, several revisions were developed and the corresponding prototypes were extensively tested. In this contribution our latest design revision proposed for the CTA MSTs are presented.
\end{abstract}

\section{INTRODUCTION}
The Cherenkov Telescope Array (CTA) is the next-generation ground-based Cherenkov telescope facility currently under preparation by an international consortium of more than 1200 scientists and engineers from 32 countries. It will be the first ground-based gamma-ray observatory open to scientists of the astronomy and particle physics communities. At two sites, one on the northern and one on the southern hemisphere, CTA will host a large number of imaging atmospheric Cherenkov telescopes of three different size classes to cover the photon energy range from a few tens of GeV to a few hundreds of TeV with a sensitivity ten times higher than currently existing arrays. This will be achieved by the combination of three different telescope classes: Large Size Telescopes (LSTs) with $\sim$23\,m, Medium Size Telescopes (MSTs) with $\sim$12\,m, and Small Size Telescopes (SSTs) with $\sim$4\,m reflector diameter. The current status of the CTA project is presented in \cite{CTA_status}.

One of the fields in which the Institute for Astronomy and Astrophysics T\"{u}bingen (IAAT) participates in CTA is the development and production of actuators for the mounting and alignment of the individual mirror segments of the MST. This contribution summarizes the current status and the future prospects of this effort.

\section{THE MIRROR CONTROL SYSTEM OF THE MEDIUM SIZE TELESCOPE}
The MSTs are dominating the sensitivity of CTA in the core energy range from 100\,GeV up to 10\,TeV. The current plan is to build 25 MSTs at the southern and 15 at the northern site as a baseline; an initial stage is planned with 15 MSTs in the south and five in the north. A detailed overview covering the various aspects of the MST design is presented in \cite{Gamma_MST}.

The effective mirror area of an MST of more than 88\,m$^2$ is reached by means of 86 hexagonal mirror segments with 1.2\,m side-to-side width. For such a large number of individual mirrors, motorized actuators are required for the initial alignment and potential realignments to achieve the optimal point spread function (PSF). Realignments might be necessary to compensate initial or long-term structure deformations, either cumulative on timescales of weeks or months or at different elevation angles and, therefore, several times per night.

\begin{figure}[t]
  \centerline{\includegraphics[width=230pt]{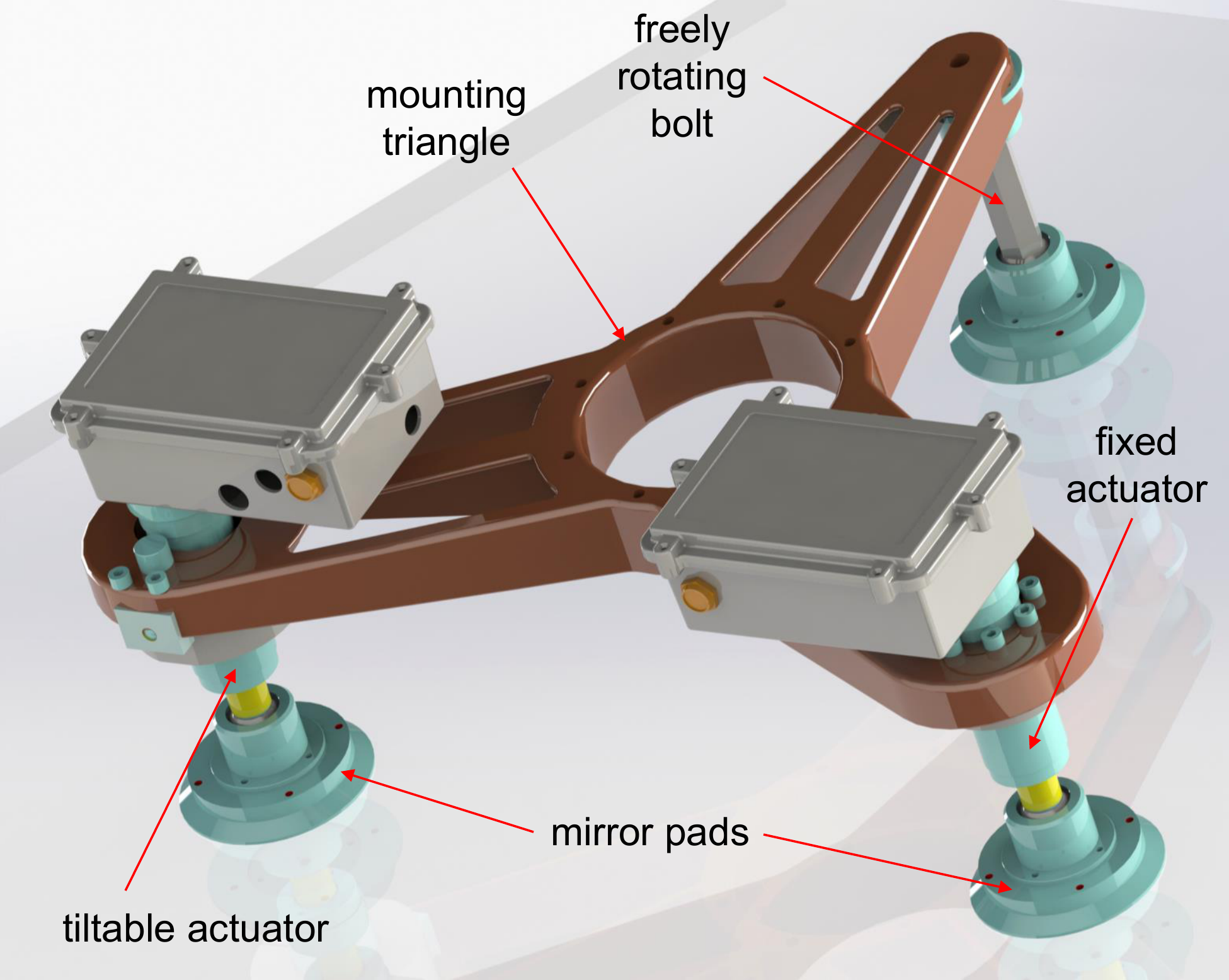}}
  \caption{The IAAT actuator concept for the CTA MST comprises one fixed and one tiltable actuator and a freely rotating bolt. The mirrors are interfaced via standardized pads; the interface to the telescope dish is a casted triangle. The triangle shown here is a preliminary design by DESY that will be modified for the production according to the final actuator design.}
  \label{fig:concept}
\end{figure}

Figure \ref{fig:concept} shows the actuator concept with one fixed and one tiltable actuator and a freely rotating bolt at the top as proposed by IAAT for the MST. The interfaces to the mirror are standardized pads that are glued on the back side of the mirror. The interface to the telescope structure is a casted triangle on which the actuators and the freely rotating bolt are mounted.

In order to achieve an optical performance, i.e. a PSF, within the MST requirements the position accuracy of the actuators should not exceed 10\,\textmu m under normal observing conditions. On the other hand, the actuators have to take the forces of a mirror with a mass of up to 40\,kg with up to 50\,kg ice on top without major damage under all environmental conditions, e.g. strong winds or large accelerations during an earthquake. Resonances in the dish are expected to scale up the loads on the mirror mounting during such events, therefore, all parts and interfaces are to be designed for loads of 5\,kN, ideally with an additional safety margin.

\section{SPECIFICATIONS OF THE LATEST IAAT ACTUATOR DESIGN ITERATION}
\begin{figure}[h]
  \centerline{\includegraphics[width=200pt]{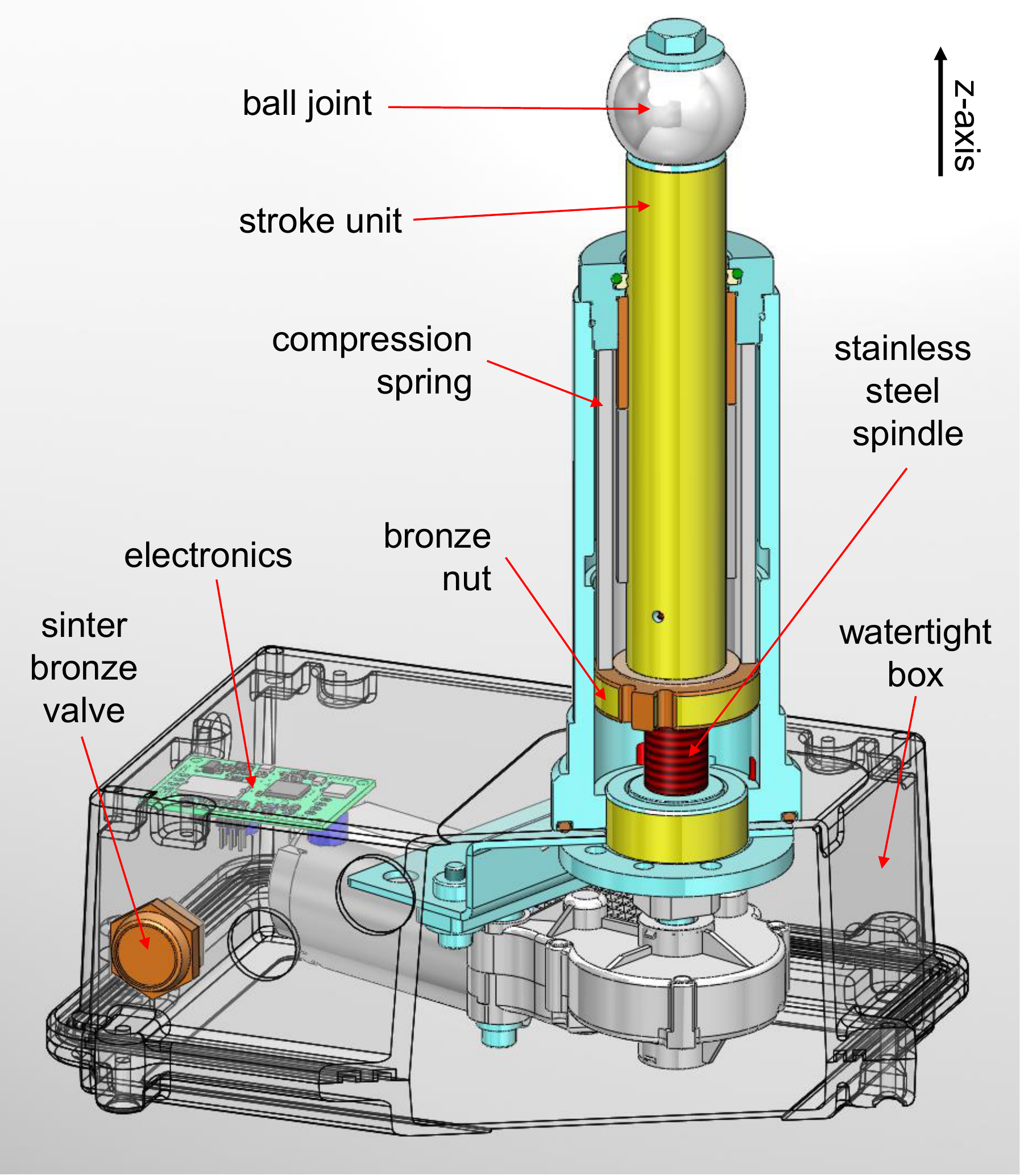}}
  \caption{The latest IAAT actuator design revision applies an internal compression spring to eliminate the mechanical play of spindle and nut and of the slide bearing that fixes the z-position of the spindle. The force of the spring is larger than the weight and wind loads at operation conditions in all positions of the stroke unit and all elevation angles of the telescope.}
  \label{fig:design}
\end{figure}

The current actuator design (cf. Fig. \ref{fig:design}) is based on the IAAT H.E.S.S. phase II heritage. It implements improvements from the experience gained at H.E.S.S. and tests conducted at IAAT and on the MST prototype in Berlin-Adlershof. The design presented here is the result of three redesign iterations in order to comply with the requirements on position accuracy as well as on earthquake-, ice-, and wind-load stability for both CTA sites. The main change in the latest iteration is the implementation of internal springs to eliminate the mechanical play of the actuator under the operation conditions defined for CTA. For the same reason additional springs have been added in the mounting of the tiltable actuator and ball joints with a minimal radial tolerance have been chosen for the actuators and for both ends of the freely rotating bolt. The production of new prototypes according to this design was started in June 2016.

The central elements of the actuator are a stainless steel spindle and a bronze nut fixed to the stroke unit. The spindle-nut combination features a thread with 2\,mm pitch and enables a Z-axis stroke of 44\,mm. The Z-axis of the spindle is fixed by means of a slide bearing to the actuator tube. The surfaces of the spindle and the tube that are in contact to the nut are hardened by plasma nitration to avoid seizing on long terms without the need for regular lubrication. For the initial greasing during production molybdenum disulfide paste is applied.

The spindle is moved by a DC motor with a gearing and an internal Hall sensor. The reading of the Hall sensor is 420 counts for a full rotation, which leads in combination with the 2\,mm pitch of the spindle to a position resolution of about 4.8\,\textmu m. This theoretical resolution is only reachable if the complete chain of mounting elements between the triangle and the mirror are play free. Therefore, an internal compression spring is used to eliminate the mechanical play between spindle and nut as well as the play in the slide bearing of the spindle. Additional play arises in the tiltable mounting of one of the actuators, which is eliminated by two springs, and in the ball joints. First tests with play-free pre-tensioned ball joints were not satisfying because they require a large torque of more than 10\,Ndm for a movement. As an alternative the use of ball joints with a quite small but finite radial tolerance is under evaluation.

The motor and the control interface are mounted in a water-tight box (IP66 certified). The box is not gas-tight in order to lower the stress on the fittings of the actuator by temperature changes. Since condensation inside the actuator might occur, a sintered bronze valve on one side of the box serves as an outlet for water vapor.

The actuators are powered with 24\,V and controlled via a CAN bus interface on custom receiver boards. The central control electronics features up to 16 mezzanine CAN controller cards, each driving an individual CAN bus, and is running a C-socket-based server to be interfaced via Ethernet.

\section{TEST RESULTS WITH EARLIER PROTOTYPES}

Long-term wear tests of the latest actuator prototypes currently available were performed with an outdoor setup at IAAT (cf. Fig. 3). The intention for these tests is to check the durability, particularly of the mechanics but also of the receiver electronics, and to verify the long-term operability under different temperature conditions. The setup emulates load on the actuators by a dummy mirror plate that is mounted in the same scheme as proposed for the MST. Details of the third design revision prototypes that were used for the tests are given in \cite{ICRC2015}.

\begin{figure}[h]
  \centerline{\includegraphics[width=\textwidth]{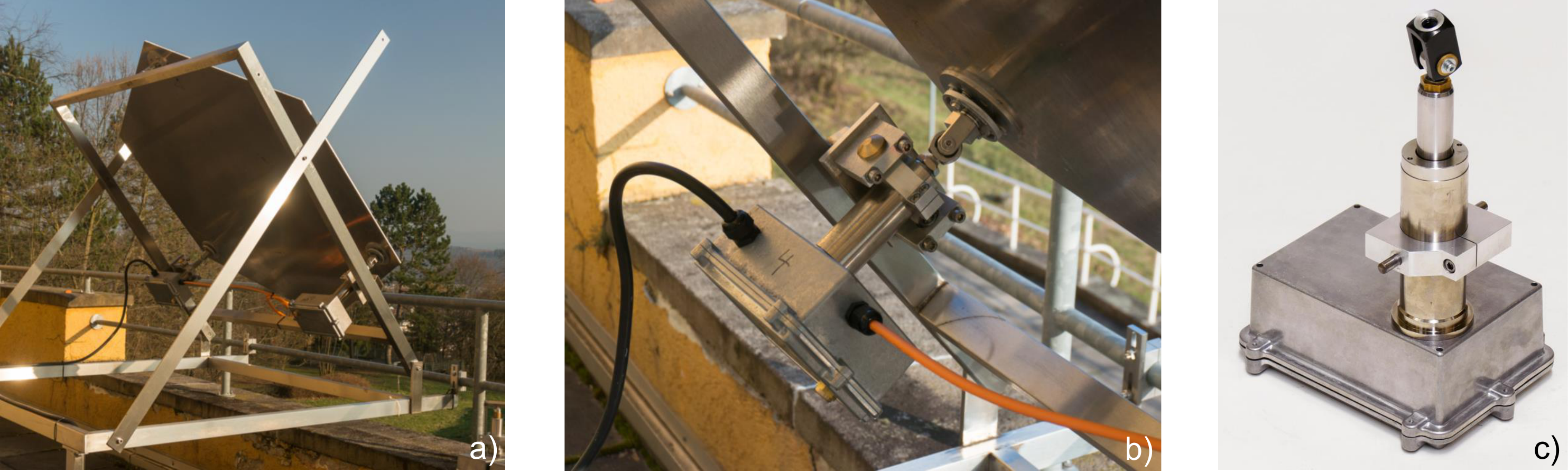}}
  \caption{Actuator prototypes are tested at IAAT in an outdoor test setup (a) that uses the same mounting scheme as proposed for the MST (b) and a dummy mirror. Prototypes of the previous design revision (c) were intensively tested in this setup and showed no degradation after long-term wear tests comparable to the estimated movement load during an MST lifetime of 30 years.}
\end{figure}

\begin{figure}[h]
  \centerline{\includegraphics[width=.9\textwidth]{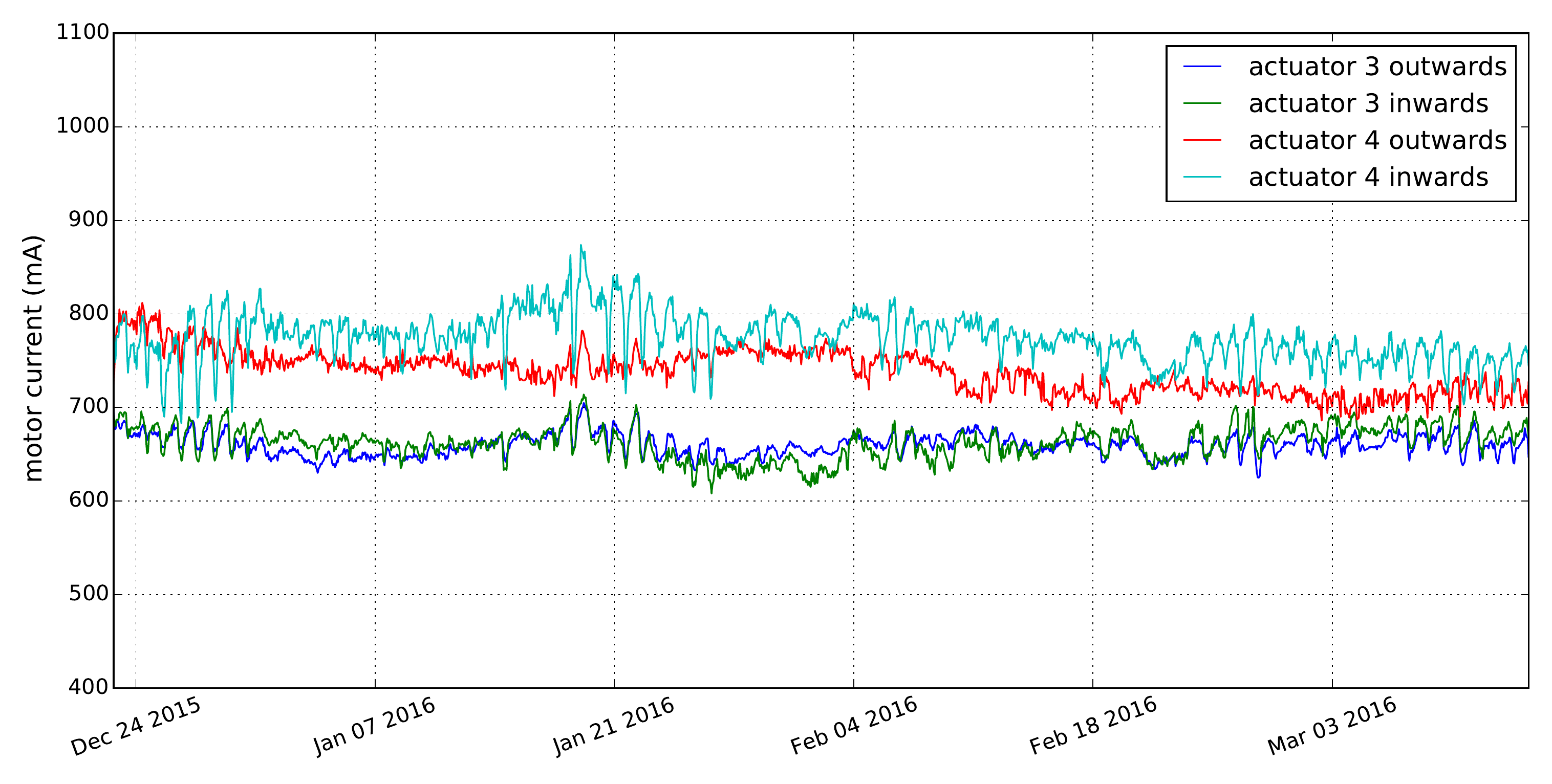}}
  \caption{Motor current monitored during long-term outdoor tests of previous revision prototypes. The actuators were continuously driven in and out over the complete movement range of 44\,mm during this measurement. The variations of the current are due to the changing ambient temperature which introduces changes of the friction. The measured current was throughout the test within the requirements and well below the turn-off threshold set in the control electronics in order to avoid any damage.}
  \label{fig:meas_full}
\end{figure}

\begin{figure}[h]
  \centerline{\includegraphics[width=.9\textwidth]{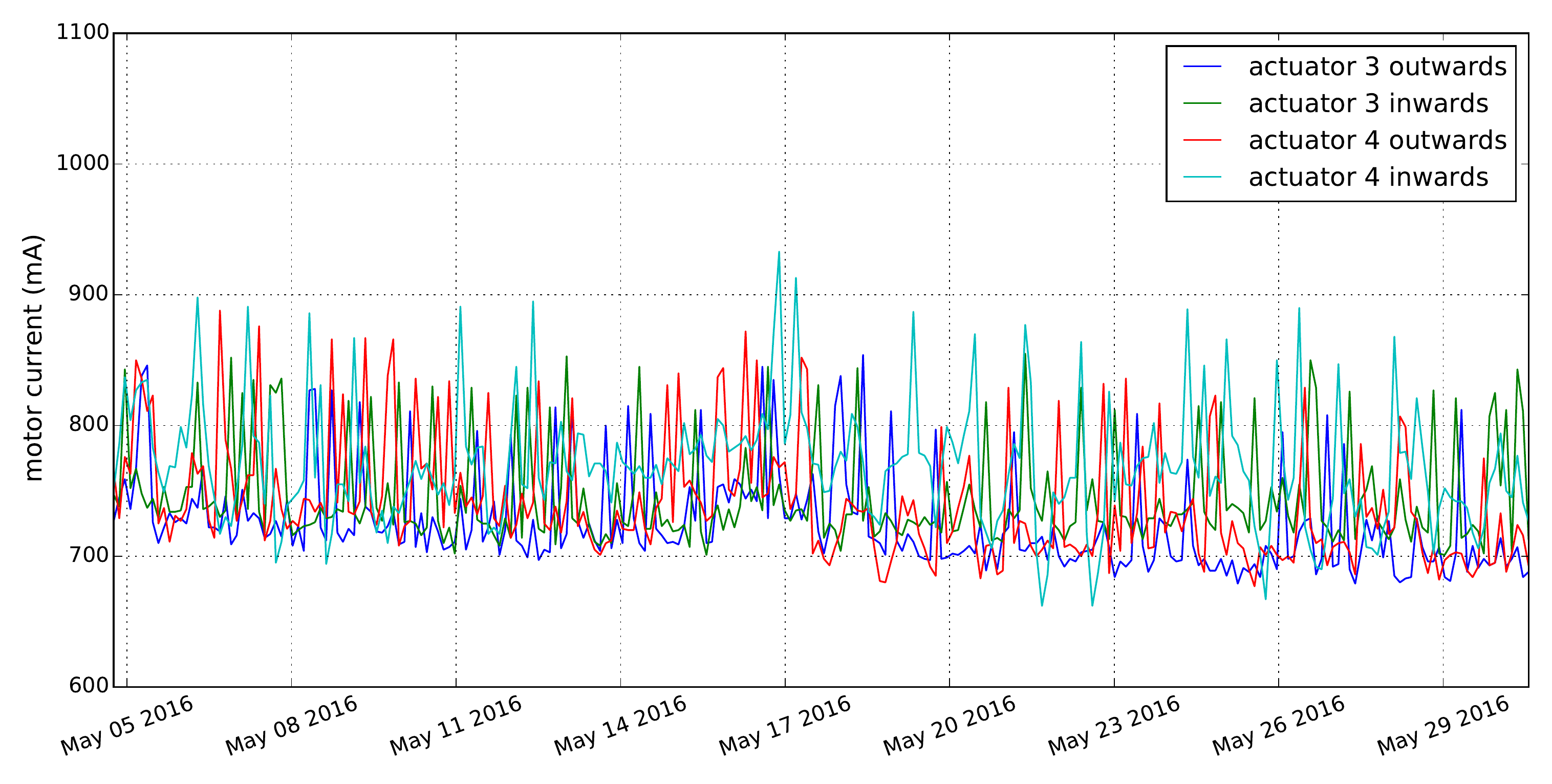}}
  \caption{Motor current monitored during long-term outdoor tests of previous revision prototypes. For this measurement the actuators were moved in and out over a range of about 1\,mm. This scenario is considered to be similar to a final use case, in which only movements over a small range are needed to realign the individual mirror spots. The spikes are due to the normal inrush current of the motor.}
  \label{fig:meas_1mm}
\end{figure}

The motor and the gearing under test are still used in the latest design and the tested spindle-nut combination is close to the latest design iteration. The only difference is that the hardening of the surface of the tested prototype was achieved by plating while for the latest design plasma nitration is applied. A test over the complete stroke range with a total movement load considerably larger than an estimated MST lifetime of 30 years revealed no measurable degradation and no blocking occurred under different temperature conditions (cf. Fig. \ref{fig:meas_full}). The same holds true for a more realistic small range wear test that probed only a 1 mm fraction of the spindle with about 100,000 up/down cycles corresponding to more than 15 years of operation with 20 movements per day (cf. Fig. \ref{fig:meas_1mm}).

\clearpage

\section{SUMMARY AND OUTLOOK}
The adaptation and optimization of the H.E.S.S. phase II actuator design to the requirements of the MST for CTA is ongoing at IAAT. Prototypes of the third revision were extensively studied and tested and the lessons learned are incorporated in a fourth revision, which is expected to be very close to the final design. The main innovation for this revision is the application of internal springs to minimize the mechanical play. Prototypes of this latest design are currently under construction and will be tested at IAAT. A new experimental setup for a precise measurement of the combined mechanical play of the actuators, the tiltable mounting at the triangle, and the ball joints at the mirror interface is currently in the design phase. Additionally, two actuator pairs are planned to be used for a long-term test on the MST prototype in Berlin-Adlershof.

\section{ACKNOWLEDGMENTS}

We gratefully acknowledge support from the agencies and organizations under Funding Agencies at http://www.cta-observatory.org.



\nocite{*}
\bibliographystyle{aipnum-cp}%
\bibliography{proc_actuator}%

\end{document}